\documentclass[twocolumn]{article}

\usepackage[margin=0.75in]{geometry}

\usepackage{cite}
\usepackage{amsmath,amssymb,amsfonts,amsthm}
\usepackage{graphicx}
\usepackage{textcomp}
\usepackage{xcolor}
\def\BibTeX{{\rm B\kern-.05em{\sc i\kern-.025em b}\kern-.08em
    T\kern-.1667em\lower.7ex\hbox{E}\kern-.125emX}}
\usepackage{algorithm}
\usepackage{tcolorbox}
\usepackage{times}
\usepackage{soul}
\usepackage{url}

\usepackage[utf8]{inputenc}

\usepackage{nicefrac}       
\usepackage{xcolor}         

\usepackage{bbold}
\usepackage{comment}
\usepackage{todonotes}

\usepackage{mathtools}
\usepackage{svg}

\usepackage{siunitx}
\usepackage{listings}

\usepackage{algpseudocode}

\usepackage{hyperref}

\usepackage{booktabs}

\usepackage{multirow}
\usepackage{booktabs}
\usepackage{arydshln}
\usepackage[T1]{fontenc}
\usepackage[utf8]{inputenc}

\newcommand{\dashedmidrule}{\noalign{\vskip 3pt}\cdashline{1-10}\noalign{\vskip 2pt}}

\title{UnWeaving the knots of GraphRAG -- turns out VectorRAG is almost enough}

\author{
 \textbf{Ryszard Tuora\textsuperscript{*}\textsuperscript{\dag}},
 \textbf{Mateusz Gali\'{n}ski\textsuperscript{*}\textsuperscript{\dag}},
 \\
  \textbf{Micha\l{} T. Godziszewski\textsuperscript{\dag}},
  \textbf{Micha\l{} Karpowicz\textsuperscript{\dag}},
 \textbf{Mateusz Czy\.{z}nikiewicz\textsuperscript{\dag}},\
 \\
  \textbf{Adam Kozakiewicz\textsuperscript{\dag}},
 \textbf{Tomasz Zi\k{e}tkiewicz\textsuperscript{\dag}}
\\
\\
 \textsuperscript{\dag}Samsung AI Warsaw
\\
 \small{
   \textsuperscript{*} Co-first authors
}
\\
\small{
   \textbf{Correspondence:} \href{mailto:m.galinski@samsung.com}{m.galinski@samsung.com},
   \href{mailto:r.tuora@samsung.com}{r.tuora@samsung.com}
 }
}

\begin{document}
\date{June 8, 2026}
\maketitle
\begin{abstract}

One of the key problems in Retrieval-augmented generation (RAG) systems is that chunk-based retrieval pipelines represent the source chunks as atomic objects, mixing the information contained within such a chunk into a single vector. These vector representations are then fundamentally treated as isolated, independent and self-sufficient, with no attempt to represent possible relations between them. Such an approach has no dedicated mechanisms for handling multi-hop questions.

Graph-based RAG systems aimed to ameliorate this problem by modeling information as knowledge-graphs, with entities represented by nodes being connected by robust relations, and forming hierarchical communities. This approach however suffers from its own issues with some of them being: orders of magnitude increased componential complexity in order to create graph-based indices, and reliance on heuristics for performing retrieval. 

We propose UnWeaver, a novel RAG framework simplifying the idea of GraphRAG. UnWeaver disentangles the contents of the documents into entities which can occur across multiple chunks using an LLM. In the retrieval process entities are used as an intermediate way of recovering original text chunks  hence preserving fidelity to the source material. We argue that entity-based decomposition yields a more distilled representation of original information, and additionally serves to reduce noise in the indexing, and generation process.

Furthermore we experimentally show that on end to end QA evaluation VectorRAG performs better than standard GraphRAG and almost as good as current SOTA graph-based solutions, for a fraction of the cost. 
\footnote{Code is available at  \href{https://github.com/SamsungLabs/UnWeaver}{\textcolor{blue}{Github}}}
\end{abstract}
\section{Introduction}\label{sec:intro}

In this paper, we propose \textbf{UnWeaver}, a novel RAG framework that achieves better than graph-like retrieval precision without explicitly building a graph. The core idea is to provide to an LLM a query-relevant context based on vector similarity of an incoming query to the embeddings formed from the concatenated descriptions of equivalent entities (or, ideas) extracted from each document chunk. Based on that similarity of the query and the distilled entity-related content, distributed over all text chunks, query-relevant chunks are easily identified with a simple index (linear mapping). Then, the retriever produces relevant context from those chunks for LLM to perform well-grounded reasoning and accurate question answering.

The core novelty here is the concatenation of chunk-based (local) descriptions of equivalent entities extracted from each chunk separately. When transformed into embeddings, these aggregated entity descriptions filter out what is irrelevant while amplifying pertinent information in relevant chunks. The results of experimental studies support that claim by showing how facts and evidence in the produced context guide LLM towards good answers and support simple reasoning. 
 \paragraph{\textbf{Related work: VectorRAG vs GraphRAG}}
Large language models (LLMs) augmented with retrieval have become a cornerstone for knowledge-intensive tasks, from question answering to fact-checking. In a Retrieval-Augmented Generation (RAG) framework, LLM is supplied with relevant external documents (or document fragments) as context, which helps ground its output in factual information. This approach has been shown to significantly reduce open-domain hallucinations by allowing the model to draw on up-to-date or domain-specific knowledge instead of relying solely on parametric memory \cite{shuster2021retrievalaugmentationreduceshallucination}. The original RAG system by Lewis et al.\cite{lewis2021retrievalaugmentedgenerationknowledgeintensivenlp}
demonstrated improved accuracy on knowledge-intensive NLP tasks by retrieving supporting passages. 

The effectiveness of a RAG-based system hinges on the quality of retrieval. Irrelevant or poorly chosen context can inject false or unrelated facts, increasing probability of factual errors or flawed reasoning in responses generated by LLM.

One of key limitations of standard RAG pipelines is their reliance on coarse-grained text chunks. Documents are typically broken into fixed-length chunks that form the units of retrieval. While this helps the retrieved (hopefully) query-related context fit within LLM input limits, in practice these retrieved parts of available text may still contain superfluous or unrelated information. 
Consequently, RAG may amplify factual inaccuracies and hallucinations in generated responses, contrary to expectations \cite{dubanowska2025representation}. Without external verification, LLMs generating response based on such a noisy context may follow off-topic details and incorporate them into their answers. Indeed, recent analyses confirm that filtering context yields noticeably more correct and concise answers \cite{joren2025sufficientcontextnewlens}.\footnote{In fact, as demonstrated by Karpowicz \cite{Karpowicz2025OnTF}, since there is no way to eliminate LLM hallucinations, we can only manage the related tradeoffs. Designing RAG systems is one of the most reasonable ways to do that, provided we have a good way of retrieving relevant knowledge (in a way aligned with our preferences).}

A promising way to address that problem is to incorporate structured knowledge graph (KG) into the RAG pipeline (see the survey \cite{graphrag}). The motivation for graph-based RAG (GraphRAG for short) is that a knowledge graph can represent information at a finer granularity. Knowledge graph essentially \emph{unweaves} the source text into a graph of connected entities (e.g., people, places, concepts), with labeled edges representing discovered relationships (their strength, nature or description).  Such graph-structured data can make it easier to reason over relevant facts without interference from unrelated context. 
In principle, the structured nature of a KG promises deeper and more contextual retrieval than flat text similarity search.

Despite their well-known advantages, GraphRAG solutions come with significant practical challenges. Constructing and maintaining a graph index of a large text corpus is a non-trivial endeavor. Unlike vector index in VectorRAG systems built directly on text embeddings, a graph index requires multiple complex steps. These include, e.g., running named entity recognition, extracting relational triples or entity links, and possibly performing clustering or ontology alignment to organize the graph. This added componential complexity means high development and computation costs. For instance, the MSGraphRAG pipeline \cite{edge2025localglobalgraphrag} involves building an entire knowledge graph from the corpus, then performing hierarchical community detection on the graph and generating summaries for each cluster---an elaborate process with many moving parts. 

Furthermore, graph retrieval itself is less straightforward than vector similarity search. One cannot simply embed a query and do $k$-NN search over nodes. Instead, tailored algorithms or heuristics are needed to traverse or query the graph structure. Prior work has explored techniques like translating natural language questions into structured graph queries (in SPARQL or Cypher), performing multi-hop subgraph searches. But these often require dedicated graph database and careful hand-tuning of query logic. As a result,  GraphRAG typically demands orders of magnitude more engineering effort and can suffer from scalability issues when dealing with millions of nodes or edges. 

All that raises an intriguing question. Is it possible to obtain the benefits of graph-based retrieval---improved relevance and reduced noise---without the heavy machinery of full-scale knowledge graphs?
\paragraph{\textbf{Contributions}}

\begin{figure*}[t]
    \centering
    \includegraphics[width=\textwidth]{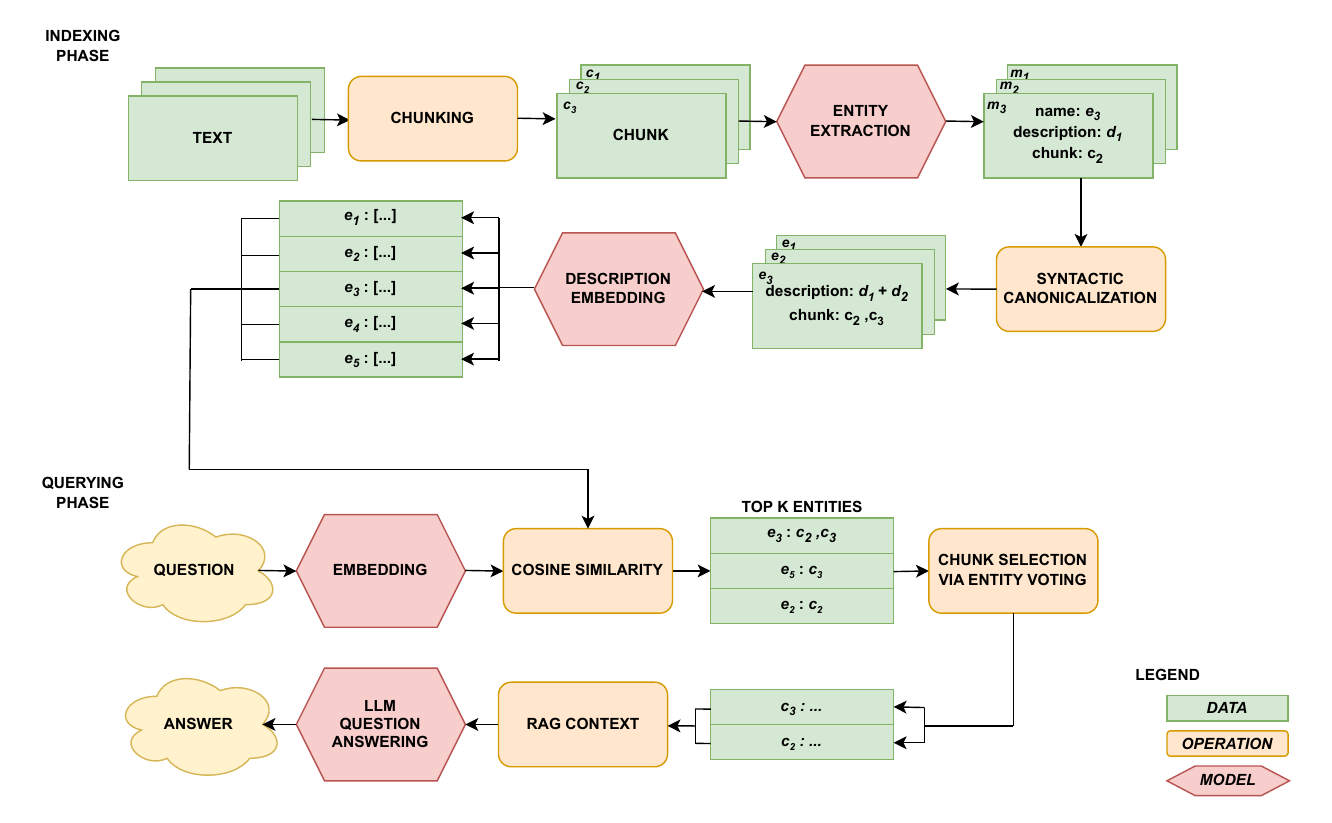}
    \caption{Architecture of UnWeaver. Model operations, shown in red, rely on an LLM, whereas operations in orange are purely algorithmic.}
    \label{fig:unweaver_flow}
\end{figure*}

The approach we propose in this paper leverages entity-centric retrieval inspired by knowledge graphs, yet it maintains the simplicity and speed of a standard vector-based pipelines (operating on refined text embeddings). That suggests there is really no need for complex and costly graph traversal or maintaining separate graph databases when documents contain separable ideas that can be named and described without much ambiguity. 

Our experimental results show that UnWeaver yields higher answer accuracy and faithfulness compared to conventional chunk retrieval baselines, demonstrating a clear reduction in hallucinated or off-target details. Moreover, this is achieved with negligible additional latency, since the entity-indexing step is done offline and query-time retrieval remains a fast embedding lookup.

UnWeaver remains firmly rooted in a standard vector-RAG framework, but enriches each indexed chunk with distilled semantic signals instead of building an explicit graph. Each stored vector therefore implicitly carries a lightweight summary of its content (e.g., key entities). At query time, these enriched embeddings bias the search toward passages that share the same salient entities, effectively filtering out irrelevant context. As a result, UnWeaver's retrieved context is far more coherent: it can mitigate data noise and reinforce logical coherence in its results, all while requiring no graph operations and preserving the efficiency of standard ANN search. In short, UnWeaver attains the benefits of GraphRAG-style structured retrieval (reduced noise, higher precision, etc.) within a pipeline that is close to pure vector-RAG.

Conceptually, UnWeaver redefines the vector index as an intelligent knowledge store, rather than a flat list of embeddings. Each stored vector is augmented with multi-level signals: alongside the raw text embedding we include a concise abstraction of the chunk's key entities (a kind of mini-graph summary). This hybrid encoding lets the retriever match on deeper features: nearest neighbors are found not just by lexical similarity, but by shared conceptual relations -- effectively capturing the intra- and inter-document connectivity that arises from explicit entity-relation structure. Crucially, these semantic gains come at no cost in scalability: the same ANN infrastructure is used, with only a lightweight preprocessing step to extract entity/relation features. In summary, UnWeaver can be thought of as not merely a compromise between GraphRAG and VectorRAG, but a principled extension of the VectorRAG paradigm that amplifies salient knowledge signals within a simple vector-retrieval framework. Furthermore, the addition of entity-based index to VectorRAG framework allows for better explainability of results since entities/ideas are inherently more explainable than raw passages of text that are returned indiscriminately by typical VectorRAG.

So, we make the following contributions in this paper.

\noindent \textbf{UnWeaver Framework:} We introduce UnWeaver, a novel RAG architecture that eradicates non-essential information from candidate retrieval chunks and focuses on salient, query-relevant content. To our knowledge, this is one of the first frameworks to successfully combine entity-centric indexing with traditional vector retrieval for RAG. This architecture can be seen on Figure \ref{fig:unweaver_flow}.

\noindent \textbf{Entity-Based Indexing Pipeline:} We develop an effective pipeline for building an entity-based index from unstructured text. This pipeline captures the core factual elements of documents (akin to a knowledge graph) without requiring full graph construction, thereby combining the precision and explainability of graph-based retrieval with the simplicity of standard dense retrieval.

\noindent \textbf{Improved Retrieval Fidelity:} Through comprehensive experiments on knowledge-intensive QA benchmarks, we demonstrate that UnWeaver significantly reduces retrieval noise and improves answer fidelity compared to standard GraphRAG approaches, both on single-hop and multi-hop questions. It outperforms baseline chunk retrieval methods in grounding LLM outputs to the correct sources, which in turn mitigates hallucinations and increases factual accuracy of the generated responses.

\noindent \textbf{Practical Insights for RAG:} We analyze the trade-offs between graph-augmented and purely text-based RAG, and discuss why UnWeaver's strategy can achieve the best of both worlds. Our findings suggest that many benefits of GraphRAG can be attained with far less complexity. This has important implications for designing the next generation of efficient and trustworthy RAG systems, indicating that graphs are not what you need if your goal is high-quality retrieval augmentation without undue overhead.

\section{Mathematics of UnWeaver}

We begin with an informal description of the algorithm. The unstructured data are passed into a (possibly multimodal) generative language model after segmentation into chunks. The language model produces a structured output listing entities mentioned within the chunk along with their descriptions. An LLM outputs across the entire document base are then aggregated, to recognize recurring entities, while retaining links to their original sources (chunks). Each entity is assigned a vector by the embedder, based on the textual representation of its features, and these are inserted into the vector store. During retrieval, the query is compared to the entity vector store, to recover the most relevant entities. The links to original document sources are then recovered and aggregated (e.g. by means of voting), to produce the most relevant document chunks. The retrieved document chunks are passed into the generator LLM, to provide the final answer.

\paragraph{\textbf{Formalized Characterization}}

Assume we are given a text $T$ (or a set of texts, documents, etc.) that will serve as the data for our RAG task. Let $C_T = (c_1, \ldots c_n)$ be the  list of chunks that the aforementioned text $T$ is divided into by some tokenizer -- i.e., $C_T$ is simply the segmented version of our text $T$. That is the first (trivial) step of our framework.

In the second step, an LLM generates names of entities together with their descriptions, from the chunks, the result of which is $n$ lists $L_1, \dots, L_n$ of pairs in the following form -- for each $i \leq n$: $L_i = (e_1, \ldots, e_{k_i})$, where for each $j \leq k_n$ the entity $e_j$ has the form $e_j = (n_j^i, d_j^i)$, where $n_j^i$ is a name of the entity, and $d_j^i$ is the description of the entity, generated by the LLM from the chunk $c_i$ (and, possibly, its internal parameters). We then glue the entities into equivalence classes of the relation of syntactic equivalence. In other words, we identify the entities $e_a= (n_a, d_a)$ and $e_b=(n_b, d_b)$, if the corresponding names  $n_a$ and $n_b$ are syntactically equivalent.

Next, given an entity $e_a$, and its equivalence class $[e_a]_{\approx_{name}}=\{e_{a_1}, \ldots, e_{a_k}\}$,\footnote{Without loss of generality, we may assume that the numbering $a_1, \ldots, a_k$ is an increasing subsequence of the ordering $1, \ldots, n$, i.e., that it is in accordance with the numbering of the chunks.} we define its representative $\tilde{e}_a$ as the following pair: $\tilde{e}_a := (n_a, \tilde{d}_a),$
where $\tilde{d}_a = {d_{a_1}}^{\frown} {d_{a_2}}^{\frown}\ldots^{\frown}d_{a_k}$ is the concatenation of the descriptions of the entity $e_a$ (possibly from different chunks)\footnote{The descriptions might be shortened by an LLM if the surpass some length threshold. This is important from the point of view of scalability.}. 

We can describe the details of the procedure, by which the choice of the representatives is being done, using the language of matrix equations. This way, we get a more fine-grained characterization of the mathematical mechanics of the algorithm. 

We embed these elements of the quotient-list of entities $\tilde{E}=\{\tilde{e}^1_1, \ldots, \tilde{e}^1_{k'_1}, \tilde{e}_1^2, \ldots, \tilde{e}_{k'_2}^2, \ldots, \tilde{e}_1^n, \ldots, \tilde{e}_{k'_n}^n\},$
into some $l$-dimensional vector space, using a specified embedder $\mathcal{E}$. For simplicity, let's use $m$ as the number of elements of the list $\tilde{E}$. This way, we obtain an $m$-element list of $l$-dimensional vectors $\mathcal{E}(\tilde{E})=(v_1, \ldots, v_m)$, where for each $i \leq m$ we have $v_i = \mathcal{E}(e_i)$. 

Additionally, we define an $m \times n$-dimensional binary matrix $M_C = [t_{ij}]_{1\leq i \leq m; 1 \leq j \leq n}$, where each row corresponds to an entity $\tilde{e}_i$, and each column corresponds to a chunk $c_j$, and $[t_{ij}] = 1$ iff the name $n_i$ has been generated from the chunk $c_j$, that is: $t_{ij} = 1$ if $n_i$ appears in $c_j$, and 0 otherwise.
Obviously, there can be more than one 1s in $i$-the verse of the matrix, if the name $n_i$ occurs in more than one chunk. 

Now, we perform the crucial step of the UnWeaver: given a query $q$, using a chosen similarity function $\sigma$, for each vector $v_i$, $i \leq m$, we compute the similarity $\sigma(v_i, \mathcal{E}(q))$ between the vector $\mathcal{E}(e_i)$, and the embedding of the query $q$.
Now, given some $k_0$, we compute the $k_0$ vectors $v_i$ with the highest similarity scores  $\sigma(v_i, \mathcal{E}(q))$, and we put $q_i:=1$ if $v_i$ is among the top-$k_0$ ones, otherwise we put $q_i:=0$, 
and we  refer to the binary vector consisting of entries $q_i$ as $M_q$. 
Then we multiply each $i$-th row of $M_C$ by the $i$-th entry of $M_q$\footnote{In other words, we pick the $i$-th row of $M_q$, which happens to be a vector of length 1, so without loss of generality we use the only scalar that is in this row.}, effectively filtering out the rows $i$ of the matrix $M_C$ which correspond to entities that are not semantically similar enough to the embedding of the query, that is we consider the intermediate filtering matrix:
$$
\left[\begin{smallmatrix}
q_1t_{11} & q_1t_{12} & \ldots & q_1t_{1n} \\
q_2t_{21} & q_2t_{22} & \ldots & q_2t_{2n} \\
\ldots & & \ldots \\
q_mt_{m1} & q_mt_{m2} & \ldots & q_mt_{mn}
\end{smallmatrix}\right],
$$ 

and we delete the rows $i$ for which $q_i=0$, so that we obtain the modified matrix $M'_C$, which is a final filtered version of the binary matrix $M_C$ with the same number of columns $n$, yet possibly a (much) smaller number of non-zero rows $m'$:
$$
M'_C= \left[\begin{smallmatrix}
q_{1'}t_{1'1} & q_{1'}t_{1'2} & \ldots & q_{1'}t_{1'n} \\
q_{2'}t_{2'1} & q_{2'}t_{2'2} & \ldots & q_{2'}t_{2'n} \\
\ldots & & \ldots \\
q_{m'}t_{m'1} & q_{m'}t_{m'2} & \ldots & q_{m'}t_{m'n}
\end{smallmatrix}\right]
$$
$$= \left[\begin{smallmatrix}
t_{1'1} & t_{1'2} & \ldots & t_{1'n} \\
t_{2'1} & t_{2'2} & \ldots & t_{2'n} \\
\ldots & & \ldots \\
t_{m'1} & t_{m'2} & \ldots & t_{m'n}
\end{smallmatrix}\right],
$$ 
where the last equality holds, since all the $q_{i'}$s are equal to 1. Finally, after depriving the non-sufficiently similar entities of their voting  rights, we perform a multi-winner approval election, where the set of remaining (non-filtered) entities (i.e., the entities $\tilde{e}_i$, for which $q_i=1$) are the set of voters, and the set of chunks $\{c_1, \ldots, c_n\}$ is the set of candidates. Each row of the matrix $M'_C$ corresponds to an approval ballot of the voter $i$, i.e., we postulate the the voter $i$ approves of the chunk $j$ iff $t_{i'j}=1$. We compute the election, according to some chosen committee approval-based election rule\footnote{This could be e.g., Approval, Proportional Approval, or Chamberlin-Courant.}, assuming we need to choose a set of $r$ chunks for some fixed $r <n$. The query response is then based only on the selected $r$ chunks.

\paragraph{\textbf{Benefits}}

Observe that the solution is by far not ad-hoc. Already on the conceptual level, there are many benefits of entity-based representation:

-- The conceptual model of entities is a ubiquitous element of human cognitive architecture, which supports the idea that this method of representing knowledge is efficient. Additionally, its naturality makes the functioning of the system more explainable and user-friendly.

-- The same entity can occur in multiple chunks, which allows to integrate information, facilitating multi-hop question answering. These chunks can represent different spatiotemporal states, different contexts, or possibly different states-of-knowledge, allowing for modeling dynamics. Additionally, thanks to the discriminatory nature of UnWeaver, we easily get rid of  superfluous or unrelated information contained in noisy coarse-grained chunks. 

-- Entities are naturally represented as belonging to a hierarchy of types (ontology). This allows for handling queries which require aggregation, i.e. exhaustive search over a selected domain of objects.

-- The lack of limits of the number of entities extracted per chunk makes it possible to disentangle an arbitrary number of topics from a chunk therefore solving the inherent problem of mixing topics in naive vector representation of chunks.
\section{Experimental Results}

\begin{table*}[ht]
    \centering
    \caption{QA performance of RAG methods}
    \label{table:evaluation}
    \scalebox{0.8}{
    \begin{tabular}{r l r r r r r r r r}
        \toprule
        \multirow{3}{*}{dataset} &
        \multirow{3}{*}{Method} &
        \multicolumn{2}{c}{FC} &
        \multicolumn{4}{c}{LLM Tokens} &
        \multicolumn{2}{c}{Embedder Tokens} \\
        & & & & \multicolumn{2}{c}{prompt} & \multicolumn{2}{c}{completion} & & \\
        & & $\mu \uparrow$ & $\sigma^{2} $ &
        {index $\downarrow$} &
        {query $\downarrow$} &
        {index $\downarrow$} &
        {query $\downarrow$} &
        {index $\downarrow$} &
        {query } \\
        \midrule
        \multirow{5}{*}{\shortstack[l]{COVID-QA \\ (1234 questions)}} &
        \textbf{UnWeaver}  & \textbf{0.3823} & 1.32$e^{-5}$ & 1\,466k & 994k & 4\,043k & \textbf{230k} & 2\,224k & 17k \\
        & \textbf{UnWeaverPPR}  & 0.3701 & 4.37$e^{-6}$ & 1\,466k & 1\,173k & 4\,043k & 285k & 2\,224k & 17k \\
        & VectorRAG & 0.3790 & 6.02$e^{-6}$ & - & \textbf{922k} & - & 359k & \textbf{476k} & 17k \\
        & RAPTOR    &  0.3643 & 8.04$e^{-6}$ & \textbf{770k} & 2\,649k & \textbf{389k} & 305k & 730k & 17k \\
        & GraphRAG  & 0.3092 & 3.66$e^{-6}$ & 25\,882k & 7\,051k & 14\,239k & 282k & 20\,455k & 17k \\
        & HippoRAG2 & 0.3735 & 6.06$e^{-6}$ & 7\,520k & 5\,260k & 7\,658k & 646k & 2\,076k & 17k \\
        \dashedmidrule
        & ClosedBook & 0.2135 & 1.50$e^{-6}$ & - & 153k & - & 283k & - & 17k \\
        & Oracle & 0.4765 & 3.77$e^{-6}$ & - & 765k & - & 258k & - & 17k \\
        \midrule
        
        \multirow{4}{*}{\shortstack[l]{eManual \\(522 questions)}} & \textbf{UnWeaver}  & \textbf{0.4811} & 2.47$e^{-5}$ & 98k & 940k & 198k & 96k & 105k & 5.7k \\
        & \textbf{UnWeaverPPR}  & 0.4514 & 1.39$e^{-5}$ & 98k & 1\,072k & 198k & 106k & 105k & 5.7k \\
        & VectorRAG & 0.4793 & 2.16$e^{-5}$ &  - & \textbf{343k} & - & \textbf{83k} & \textbf{36k} & 5.7k \\
        & RAPTOR    &  0.4466 & 2.17$e^{-5}$ & \textbf{57k} & 1\,177k & \textbf{30k} & 115k & 60k & 5.7k \\
        & GraphRAG  & 0.3424 & 5.87$e^{-6}$ & 1\,428k & 3\,266k & 687k & 112k & 1\,063k & 5.7k \\
        & HippoRAG2 & 0.4680 & 1.43$e^{-5}$ & 468k & 2\,500k & 356k & 278k & 104k & 5.7k \\
        \dashedmidrule
        & ClosedBook & 0.1793 & 5.01$e^{-6}$ & - & 64k & - & 105k & - & 5.7k \\
        & Oracle & 0.5318 & 2.30$e^{-6}$ & - & 345k & - & 90k & - & 5.7k \\
       \midrule
        
        \multirow{4}{*}{\shortstack[l]{Tech-QA \\(596 questions)}} &
        \textbf{UnWeaver}  & 0.3135 & 7.54$e^{-6}$ & \textbf{3\,747k} & 2\,384k & \textbf{4\,446k} & \textbf{158k} & 2\,411k & 55k \\
        & \textbf{UnWeaverPPR}  & 0.3123 & 2.53$e^{-6}$ & \textbf{3\,747k} & \textbf{2\,140k} & \textbf{4\,446k} & 194k & 2\,411k & 55k \\
        & VectorRAG & 0.3218 & 7.86$e^{-6}$ & - & 2\,354k & - & 161k & 3\,026k &  55k \\
        & RAPTOR    & - & - & - & - & - & - & - & - \\
        & GraphRAG  & 0.2847 & 4.19$e^{-6}$ & 49\,270k & 5\,944k & 35\,704k & 163k & 54\,016k & 55k \\
        & HippoRAG2 & \textbf{0.3448} & 1.34$e^{-5}$ & 7\,814k & 5\,690k & 6\,258k & 564k & \textbf{4\,448k} & 55k \\
        \dashedmidrule
        & ClosedBook & 0.2052 & 4.20$e^{-6}$ & - & 118k & - & 218k & - & 55k \\
        & Oracle & 0.3367 & 1.66$e^{-6}$ & - & 2\,996k & - & 195k & - & 55k \\
        \midrule

        \multirow{4}{*}{\shortstack[l]{MuSiQue \\(2417 questions)}} &
        \textbf{UnWeaver}  & 0.2028 & 4.60$e^{-6}$ & \textbf{985k} &1\,861k & 2\,678k & \textbf{882k} & 1\,327k & 54k \\
        & \textbf{UnWeaverPPR}  & 0.1946 & 3.88$e^{-6}$ & \textbf{985k} &2\,320k & 2\,678k & 986k & 1\,327k & 54k \\
        & VectorRAG & 0.2007 & 1.47$e^{-6}$ & - & \textbf{1\,851k} & - & 883k & \textbf{285k} &  54k \\
        & RAPTOR    & 0.1858 & 1.29$e^{-6}$ & 440k & 6\,024k & \textbf{915k} & 1\,150k & 450k & 54k \\
        & GraphRAG  & 0.1353 & 7.65$e^{-6}$ & 41\,002k & 18\,943k & 27\,046k & 1\,137k & 10\,330k & 54k \\
        & HippoRAG2 & \textbf{0.2196} & 1.01$e^{-6}$ & 2\,616k & 10\,431k & 2\,723k & 1\,778k & 1\,070k & 54k \\
        \dashedmidrule
        & ClosedBook & 0.1476 & 9.82$e^{-6}$ & - & 321k & - & 1\,641k & - & 54k \\
        & Oracle & 0.2176 & 2.02$e^{-6}$ & - & 1\,097k & - & 664k & - & 54k \\
        \bottomrule
        
    \end{tabular}
    }
\end{table*}

 We evaluate the quality of RAG systems on downstream QA task with a selection of three datasets included in the distantly supervised RAGBench suite \cite{friel2025ragbenchexplainablebenchmarkretrievalaugmented}. Our choice of datasets consist of COVID-QA, eManual and Tech-QA in order to cover different, single-hop, QA scenarios. To that we add MuSiQue \cite{trivedi2022musiquemultihopquestionssinglehop} in order to cover multi-hop questions.
 
 To identify the pitfalls and / or inefficiencies of graph-based RAG systems we compare our method against Microsoft's GraphRAG\cite{graphrag}, HippoRAG2\cite{gutiérrez2025ragmemorynonparametriccontinual} as well as RAPTOR\cite{sarthi2024raptorrecursiveabstractiveprocessing}.
 
 GraphRAG was the system that sparked the domain of graph-based RAG systems, it performs a multi step indexing procedure that first extracts entities and relations and then builds a hierarchical communities using those entities. During retrieval it relies on heuristics to select communities, relations and entities in order to perform QA. 

  HippoRAG\cite{gutiérrez2025hipporagneurobiologicallyinspiredlongterm} as well as its successor HippoRAG2 are often on the forefront of the RAG systems SOTA. The successor constructs graph similar to how UnWeaver does it, with the difference that it creates entity-entity edges, which UnWeaver forgoes. At retrieval time it searches the created graph using Personalized PageRank using the top-k most similar entities to the query as the seed as well as variety of other heuristics. 
 
 Recently RAPTOR emerged and has been one of the dominant systems in graph-based RAGs in terms of its performance. It builds its index by recursively clustering summaries of chunks, it then makes use of that tree to perform QA. 
 
 The last system we compare against as baseline is VectorRAG\cite{lewis2021retrievalaugmentedgenerationknowledgeintensivenlp} that created the domain of RAG systems in the first place.
 
 All the methods are indexed and queried with \texttt{GPT-OSS-120B} LLM and embedded with \texttt{Qwen3-Embedding-4B}.  To evaluate the quality of QA we employ calculation of Factual Correctness (F1) from RAGAS\cite{es2025ragasautomatedevaluationretrieval}, with the judge also being \texttt{GPT-OSS-120B}. FC F1 is calculated from FC recall and FC precision which in turn are calculated by performing claim decomposition, via LLM, on both the ground truth answer as well as the system's answer. To account for inherent stochasticity of LLMs we perform 5 evaluation runs in order to get mean and variance of the mean FC for each method. To eliminate the effect of response formatting on FC score we fix the query prompt across all the methods to be the same. Finally we report the results of closed book as well as oracle querying in order to show the theoretical minimum and maximum achievable in this metric. The results are listed in \autoref{table:evaluation}.
 
In order to decouple the knowledge representation part from the retrieval part of UnWeaver, we've tested a modification of UnWeaver that utilities Personalised PageRank at retrieval time instead of using weighted voting via identified entities.

Across all datasets the top 3 ranking of methods always involve HippoRAG2, VectorRAG and UnWeaver, with variance in individual placements in that ranking. On the other end of the spectrum is GraphRAG which always occupies the last spot. On COVID-QA and eManual UnWeaver is the best system, while on Tech-QA\footnote{On Tech-QA dataset, we were unable to produce results using RAPTOR despite numerous attempts. The problem was numerical instability in calculation of Gaussian Mixture Models that are the backbone of RAPTOR's clustering procedure.} and MuSiQue it's HippoRAG2. 

The surprising fact is that VectorRAG is the main competitor across these datasets, with other Graph-based approaches lagging behind in terms of Factual Correctness. This can be partially explained by the architectural similarity of VectorRAG to the procedure for generating the original datasets. 

On Tech-QA and MuSiQue we can observe that not relying on source material, like VectorRAG, UnWeaver or Oracle do, and instead using transformed output, like HippoRAG2, GraphRAG and RAPTOR do, can yield good results in terms of QA. However on COVID-QA and eManual we can observe the exact opposite, showcasing the vulnerability of evaluation of RAG systems based on the type of dataset used. Although relying on too heavily transformed output, which GraphRAG is infamous for, might have devastating consequences as it can be seen by its FC being lower than that of pure LLM without context provided to it, so the ClosedBook one. 
 
 The quality of answers should be also evaluated in tandem with Token Usage to estimate the cost and latency of indexing and querying the data. With the exception for eManual, UnWeaver uses nearly the same amount of tokens in the query phase as VectorRAG, which is the most efficient system, thereby providing the lowest query latency. Furthermore the main graph-based competitor of UnWeaver, HippoRAG2, uses anywhere from $\sim1.5\times$ to $\sim2.75\times$ as many tokens during indexing. This makes UnWeaver cheaper to deploy in production setting since it manages to achieve reduction both in prompt as well as completion tokens\footnote{With the exception for MuSiQue where both UnWeaver and HippoRAG2 consumed around the same amount of completion tokens.} compared to HippoRAG2.

\bibliography{main}
\bibliographystyle{plain} 

\clearpage

\onecolumn
\appendix
\section{Mathematics of UnWeaver and Beyond}

Let $\mathbf{T} = [\mathbf{T}_1| \cdots| \mathbf{T}_K]$ be a document decomposed into $K$ chunks by a tokenizer.  From each chunk $i = 1,\dots,K$ an LLM extracts a list of $n$ names of entities (entities for short) together with their descriptions. Each entity description is based on the $i$-th chunk content only.

We need named entities to produce equivalence classes of the relation of syntactic (or semantic, in general case)\footnote{Involving semantic identity summons the extremely nontrivial problem of canonicalization, which is irrelevant to the main focus of the paper. In any case, the proposed mechanism applies to semantic equivalence classes perfectly without any modifications.} identity. In other words, two entities are the same, if their names are equivalent. We represent each identified equivalence class $s = 1,\dots,S$ as a permutation submatrix $\mathbf{P}_s$ selecting relevant descriptions and chunks from the two additional matrices we introduce.

First, matrix
\begin{align*}
\mathbf{A} = 
[\mathbf{a}_{11}|\cdots|\mathbf{a}_{1k(1)}|\cdots|\mathbf{a}_{n1}|\cdots|\mathbf{a}_{nk(n)}]
\end{align*}
contains descriptions of each entity represented as column vectors, i.e., $\mathbf{a}_{ij}$ is a description (string general encoding) of entity $i$ generated by LLM based on the content of chunk $j$ of the original document $\mathbf{T}$.

Then, second, the chunk-entity matrix (in general form),
\begin{align*}
\mathbf{W} = 
\begin{bmatrix}
1 & \cdots & 1 & & & & & & \\
& & & 1 & \cdots & 1 &  &  &  \\
& & & & & & 1 & \cdots & 1
\end{bmatrix}_{K\times n},
\end{align*}
encodes presence of each identified entity in each chunk of document $\mathbf{T}$ . So, $w_{ij} = 1$ is entity $j$ has been identified in chunk $i$, and $w_{ij} = 0$ otherwise. That matrix serves as a tool for relevant chunk identification. 

Given these two matrices, we can now construct operations extracting descriptions and related document chunks for each equivalence class of semantically identical entities. Let $\mathbf{I}_n$ be an $n$ by $n$ identity matrix. For an equivalence class $s$, let $\sigma(s) \subseteq \{1,\dots, n\}$ be a subset of indices of identical (equivalent) entities. Then each equivalence class is represented by the following permutation submatrix:
\begin{align*}
\mathbf{P}_s = \mathbf{I}_n(:,\sigma(s)).
\end{align*}
The desired extraction operations can now be implemented (in an elegant way) as matrix multiplications. For each equivalence class $s$, or an \textit{idea} represented by that class, the relevant descriptions are concatenated (as columns) of matrix
\begin{align*}
\mathbf{D}_s = \mathbf{A}\mathbf{P}_s 
= \mathbf{A}(:,\sigma(s)).
\end{align*}
Similarly, the indices of relevant document chunks are given by a binary vector
\begin{align*}
\mathbf{c}_s = \mathrm{sign}(\mathbf{W}\mathbf{P}_s\mathbf{1}_{|\sigma(s)|\times 1}).
\end{align*}
Then, the chunk by equivalence class (chunk-class) matrix is
\begin{align*}
\mathbf{C} = [\mathbf{c}_1|\cdots|\mathbf{c}_S]_{K\times S}.
\end{align*}

Next, we design a vector database (VDB), represented by matrix $\mathbf{V}$, by embedding the concatenated descriptions of each equivalence class:
\begin{align*}
\mathbf{V} = 
\begin{bmatrix}
\mathcal{E}(\mathbf{D}_1) |\cdots | 
\mathcal{E}(\mathbf{D}_S) 
\end{bmatrix}_{P\times S}
=
\begin{bmatrix}
\mathcal{E}(\mathbf{A}\mathbf{P}_1) |\cdots | 
\mathcal{E}(\mathbf{A}\mathbf{P}_S) 
\end{bmatrix}_{P\times S}.
\end{align*}
With that database we are now ready to design the retrieval operation. 

\section{Retrieval}

Suppose we are given a query vector $\mathbf{q}$ for which we need to prepare the context for response generation. That context must contain the retrieved relevant document chunks. The idea behind the UnWeaver-based retrieval exploits properties of the vector database $\mathbf{V}$ to identify those chunks that are relevant to the query.

For the sake of illustration, let us take a look at simple models the VDB search. First, consider the orthogonal projection of $\mathbf{q}$ onto the column-space of $\mathbf{V}$. That projection defines a vector in the column-space of $\mathbf{V}$ that is closest to the query. It is given by
\begin{align*}
\hat{\mathbf{q}} = \mathbf{V}\mathbf{V}^+\mathbf{q},
\end{align*}
where $\mathbf{V}^+$ is the Moore-Penrose pseudoinverse. Notice that to find a list of $k$ embeddings of the extracted entity descriptions that are nearest to $\mathbf{q}$, it is in fact enough to calculate the Euclidean distance
\begin{align*}
\|\mathbf{V}(:,s)-\mathbf{q}\|^2_2=
\|\mathbf{V}(:,s)-(\hat{\mathbf{q}}+\mathbf{e})\|^2_2 =
\|\mathbf{V}(:,s)-\hat{\mathbf{q}}\|_2^2+\|\mathbf{e}\|_2^2,
\end{align*}
which follows from the Pythagorean theorem for orthogonal vectors $\mathbf{V}(:,s)-\hat{\mathbf{q}}$ and $\mathbf{e}$ (projection error). Then, the relevant database entries are index by 
\begin{align*}
\hat\sigma = \mathrm{Arg}_k\min(k)_{s = 1,\dots,S}
(\|\mathbf{V}(:,s)-{\mathbf{q}}\|)_{s=1}^S.
\end{align*}
We write $\min(k)$ (or $\max(k)$) to denote $k$ best ($\min$ or $\max$) elements of the list of numbers that follows.

The relevant context is now given by retrieval matrix
\begin{align*}
\mathbf{R}=\mathbf{C}(:,\hat\sigma)_{K\times k},
\end{align*}
a submatrix built from columns $\hat{\sigma}$ in of martix $\mathbf{C}$.

Similarly, for the search based on cosine similarity, we have
\begin{align*}
\tilde{\sigma} = \mathrm{Arg} \max(k)_{s = 1,\dots,S}
(\tilde{\mathbf{V}}(:,s)^T\tilde{\mathbf{q}})_{s=1}^S \\ \text{for}\quad 
\tilde{\mathbf{V}}(:,s) = \dfrac{\mathbf{V}(:,s)}{\|\mathbf{V}(:,s)\|}
\ \text{and}\ \tilde{\mathbf{q}} = \dfrac{\mathbf{q}}{\|\mathbf{q}\|}.
\end{align*}
Here, the relevant context is also given by 
\begin{align*}
\mathbf{R}=\mathbf{C}(:,\tilde\sigma)_{K\times k}.
\end{align*}
The retrieval matrix is a binary matrix such that $r_{ij} = 1$ is chunk $i$ contains description $j$ that is one of top-$k$ descriptions of ideas (entities of equivalence class) relevant to the query. Notice, that the same chunk may contain many such descriptions, an observation motivating our next step going beyond basic retrieval.

\section{Alignment}

The retrieval framework we have developed provides an interesting setting for retrieval alignment control. Notice that matrix $\mathbf{C}$, or its retrieval submatrix $\mathbf{R}$, can be interpreted as a routing matrix in the following well studied aggregate utility maximization problem:
\begin{align*}
\text{maximize}\ &\sum_{s=1}^S U_s(x_s)\ \ 
\text{s.t.}\ \mathbf{R}\mathbf{x}\le \mathbf f \
\text{and (optionally)}\ 
-\mathbf{I}_S\mathbf{x}\le \mathbf{0}_{S\times 1}.
\end{align*}
The related Lagrange function is then
\begin{align*}
\mathcal{L}(\mathbf{x},\boldsymbol{\lambda},\boldsymbol{\mu}) = 
\sum_{s=1}^S U_s(x_s) - \boldsymbol{\lambda}^T(\mathbf{R}\mathbf{x} - \mathbf f )+\boldsymbol{\mu}^T\mathbf{x}.
\end{align*}
For a detailed study of the solutions to the problem, see e.g. \cite{Karpowicz12:AMCS,Karpowicz2012OnTD}. 

We want to find an optimal distribution $\bar{\mathbf{x}} = [x_1,\cdots,x_S]^T$ of the strength of ideas (representatives of equivalence classes identified in $\mathbf{T}$) over the relevant document chunks (containing those ideas). However, the strength of ideas in each chunk cannot exceed an imposed upper bound $\mathbf{f}$ that controls the distribution. That strength, defined by a given $\mathbf{x}$, is described by the product 
\newcommand*{\horzbar}{\rule[.5ex]{2.5ex}{0.5pt}}
\begin{align*}
\mathbf{R}\mathbf{x}=
\begin{bmatrix}
\mathbf{r}^*_1
\\
\horzbar
\\
\vdots
\\
\horzbar
\\
\mathbf{r}_K^*
\end{bmatrix}    
\mathbf{x},
\end{align*}
where $\mathbf{r}^*_k$ denotes row $k$ of the matrix.

It follows from the KKT optimality conditions, that a solution $(\bar{\mathbf{x}},\bar{\boldsymbol{\lambda}})$ to the aligned retrival problem satisfies the following system of inequalities:
\begin{align*}
U'_s(\bar{x}_s) \le \sum_{i=1}^Kr_{is}\bar{\lambda}_i,\ \text{with equality for}\ \bar{x}_s>0,\ s = 1,\dots,S,
\end{align*}
where $\bar{\lambda}_s$ denotes Lagrange multiplier for the mixing strength constraint $\mathbf{r}^*_k\mathbf{x}\le f_k$ in chunk $k =1,\dots,K$. 

It is now clear that selecting the utility function $U_s$ for each equivalence class (or extracted idea), together with the vector $\mathbf{f}$, allows for controlling the retrieval alignment. One theoretically elegant example includes
\begin{align*}
U_s(x_s) = \gamma_s(\mathbf{q}) \log(x_s)\ \text{for}\ \gamma_s =\gamma_s(\mathbf{q}) .
\end{align*}
Here, weight $\gamma_s$ could depend on the relevance of the available database entries in $\mathbf{V}$ for given query $\mathbf{q}$ vector. For example, we could set
\begin{align*}
\boldsymbol{\gamma} = \mathbf{V}^T\mathbf{q}. 
\end{align*}
Then, optimal distribution of chunks is given by the well recognized interpretable closed-form solution (reachable in a convergent iterative auction-like process, e.g., see \cite{kelly1998rate}):
$$\bar{x}_s = \dfrac{\gamma_s}{\sum_{i=1}^Kr_{is}\bar{\lambda}_i}
= \dfrac{\mathbf{V}(:,s)^T\mathbf{q}}{\sum_{i=1}^Kr_{is}\bar{\lambda}_i}.$$
Another straightforward and elegant approach we can propose, defines retrieval alignment as a constrained least squares problem:
\begin{align*}
\mathrm{minimize}\ \dfrac{1}{2}\|\mathbf{V}\mathbf{x}-\mathbf{q}\|_2^2
\ \ \text{s.t.}\ \mathbf{C}\mathbf{x} = \mathbf{f}.
\end{align*}
If $\mathbf{f}$ belongs to the column space of $\mathbf{V}$, $\mathbf{C}$ has independent rows, and $[\mathbf{A}^T\ \mathbf{C}^T]^T$ has independent columns, then the problem has a unique and interpretable closed form solution $(\bar{\mathbf{x}},\bar{\boldsymbol{\lambda}})$ meeting the KKT linear matrix equation$
\begin{bmatrix}
\mathbf{V}^T\mathbf{V} & \mathbf{C}^T
\\
\mathbf{C} & \mathbf{0}
\end{bmatrix}
\begin{bmatrix}
\bar{\mathbf{x}}
\\
\bar{\boldsymbol{\lambda}}
\end{bmatrix}
=
\begin{bmatrix}
\mathbf{V}^T\mathbf{q}
\\
\mathbf{f}
\end{bmatrix}.
$
For the study of general and randomized solutions to that equation see e.g. \cite{boyd2018introduction,karpowicz_theory_2021,ben2003generalized}.

\begin{algorithm}[H]
\caption{Aligned Retrieval with UnWeaver}
\label{alg:aligned-rag}
\begin{algorithmic}[1]
\Require Document $\mathbf{T}$, query $q_{\text{text}}$, embedding operator $\mathcal{E}\colon \mathcal{T}^* \to \mathbb{R}^P$
\Ensure Context chunks

\State \textbf{Preprocessing:}
\State $[\mathbf{T}_1 | \cdots | \mathbf{T}_K] \gets \text{Tokenize}(\mathbf{T})$
\State Extract entities and descriptions: $\mathbf{A} \gets [\mathbf{a}_{ij}]_{i=1,\ldots,K}^{j=1,\ldots,n}$, $\mathbf{W} \in \{0,1\}^{K \times n}$
\State Identify equivalence classes: $\{\sigma(1), \ldots, \sigma(S)\}$
\For{$s = 1$ to $S$}
\Comment{Representatives of equivalent classes}
    \State $\mathbf{P}_s \gets \mathbf{I}_n(:, \sigma(s))$
    \State $\mathbf{c}_s \gets \text{sign}(\mathbf{W} \mathbf{P}_s \mathbf{1})$
    \State $\mathbf{v}_s \gets \mathcal{E}(\text{concat}(\mathbf{A}\mathbf{P}_s))$
\EndFor
\State $\mathbf{C} \gets [\mathbf{c}_1 | \cdots | \mathbf{c}_S] \in \{0,1\}^{K \times S}$
\State $\mathbf{V} \gets [\mathbf{v}_1 | \cdots | \mathbf{v}_S] \in \mathbb{R}^{P \times S}$

\State \textbf{Query Processing:}
\State $\mathbf{q} \gets \mathcal{E}(q_{\text{text}})$
\State $\boldsymbol{\gamma} \gets \mathbf{V}^T \mathbf{q}$ \Comment{Relevance scores for Retrieval Alignment}

\State \textbf{Retrieval Alginment:}
\If{method = Utility-Based}
    \State Find: $\arg\max_{\mathbf{x} \geq 0} \sum_{s=1}^S \gamma_s \log(x_s)$ s.t. $\mathbf{C}\mathbf{x} \leq \mathbf{f}$
    \State Initialize: $\boldsymbol{\lambda} \gets \mathbf{1}_K/K$
    \For{$t = 1$ to $t_{\max}$}
        \State $\mathbf{x} \gets [\gamma_s/(\mathbf{C}(:,s)^T\boldsymbol{\lambda})]_{s=1}^S$ \Comment{Primal update}
        \State $\boldsymbol{\rho} \gets \mathbf{C}\mathbf{x}$
        \State $\boldsymbol{\lambda} \gets \text{UpdatePrices}(\boldsymbol{\lambda}, \boldsymbol{\rho}, \mathbf{f})$ \Comment{Dual update}
        \If{$\|\boldsymbol{\rho} - \mathbf{f}\|_\infty < \epsilon$} \textbf{break} \EndIf
    \EndFor
\ElsIf{method = CLS}
    \State Find: $\arg\min_{\mathbf{x}} \frac{1}{2}\|\mathbf{V}\mathbf{x}-\mathbf{q}\|_2^2$ s.t. $\mathbf{C}\mathbf{x} = \mathbf{f}$
    \vspace{2mm}
    \State Solve: $\begin{bmatrix} \mathbf{V}^T\mathbf{V} & \mathbf{C}^T \\ \mathbf{C} & \mathbf{0} \end{bmatrix} \begin{bmatrix} \mathbf{x} \\ \boldsymbol{\lambda} \end{bmatrix} = \begin{bmatrix} \mathbf{V}^T\mathbf{q} \\ \mathbf{f} \end{bmatrix}$
\EndIf

\State \textbf{Context Retrieval:}
\State $\mathcal{S} \gets \text{TopK}(\mathbf{V},\mathbf{q},\mathbf{x}, k')$ \Comment{Best equivalence classes}
\State $\mathcal{I} \gets \{i : \exists s \in \mathcal{S}, c_{is} = 1\}$ \Comment{Relevant chunks}
\State \Return $\{\mathbf{T}_i\}_{i \in \mathcal{I}}$
\end{algorithmic}
\end{algorithm}

\newpage

\section{Numerical Example}

Suppose we have already decomposed document $\mathbf{T}$ into $K  = 2$ chunks, and we extracted with an LLM entities forming $S = 3$ equivalent ideas (or equivalence classes). Let us assume that the distribution of those ideas in the chunks is
\begin{align*}
\mathbf{C} = 
\begin{bmatrix}
1 & 0 & 1\\0 & 1 & 1
\end{bmatrix},
\end{align*}
and their concatenated descriptions are represented by the following matrix of embedding
\begin{align*}
\mathbf{V} = 
\begin{bmatrix}
1 & 0 & 0.5\\0 & 1 & 0.5
\end{bmatrix}.
\end{align*}
Given query $\mathbf{q} = \begin{bmatrix}1 & 1\end{bmatrix}^T$, we want to retrieve relevant chunks of text that are well aligned with our preferences for context generation. We set the budget of ideas in each chunk to $\mathbf{f} = \begin{bmatrix}1 & 1\end{bmatrix}^T$ and demand $L = 2$ to be kept in the retrieved context. 

With this simple settings, we get
\begin{align*}
\boldsymbol{\gamma} = \mathbf{V}^T\mathbf{q} = \begin{bmatrix}
 1\\1\\1   
\end{bmatrix},
\end{align*}
and the retrieval alignment problem becomes:
\begin{align*}
\mathrm{maximize}\sum_{s=1}^S \log(x_s)\ \text{s.t.}\  
\mathbf{C}\mathbf{x} = \mathbf{f}.
\end{align*}
It can be verified rather easily, that KKT conditions are satisfied for 
\begin{align*}
\bar{\mathbf{x}} = \begin{bmatrix}
\frac{2}{3}\\\frac{2}{3}\\\frac{1}{3}
\end{bmatrix}
\ \text{and}\ 
\bar{\boldsymbol{\lambda}} = \begin{bmatrix}
\frac{3}{2}\\\frac{3}{2}\\\frac{3}{2}
\end{bmatrix}.
\end{align*}
From that we conclude that the aligned retrieval generates context consisting of the relevant chunks
\begin{align*}
\bar\sigma = \{1,2\}.
\end{align*}

\newpage

\section{Prompts}
\newtcolorbox{promptbox}{
  fontupper=\footnotesize\ttfamily,
  left=4pt,right=4pt,top=0pt,bottom=0pt,
  arc=0mm,
}
This is the query prompt that was used across all methods.

\begin{promptbox}
\begin{lstlisting}
Role: System
Content:
    You are a question answering system.
    Please make sure that the answer is correct and complete. 
    At the same time avoid redundancy and irrelevant information.
    Please try to answer the question in Single Sentence.

    Do so based on the following context:
    
    {context}


Role: User
Content: 
    {question}
\end{lstlisting}
\end{promptbox}

\end{document}